# PERCEPTUAL EVALUATION OF PLAYOUT BUFFER ALGORITHM FOR ENHANCING PERCEIVED QUALITY OF VOICE TRANSMISSION OVER IP NETWORK

Yusuf Perwej[1] and Firoj Parwej[2]

[1]M.Tech, MCA, Department of Computer Science & Information System, Jazan University, Jazan , Kingdom of Saudi Arabia(KSA)

Yusufperwej@gmail.com

[2]MCA, MBA, BIT, Department of Computer Science & Information System, Jazan University, Jazan , Kingdom of Saudi Arabia(KSA)

firojparwej@gmail.com

## ABSTRACT

*Voice over Internet Protocol (VoIP) is a technology that allows you to make voice calls using a broadband Internet connection instead of a regular (or analog) phone line. Voice over Internet Protocol (VoIP) has led human speech to a new level, where conversation across continents can be much cheaper & faster. However, as IP networks are not designed for real-time applications, the network impairments such as packet loss, jitter and delay have a severe impact on speech quality. The playout buffer at the receiver side is used to compensate jitter at a trade-off of delay and loss. We found the characteristics of delay and loss are dependent on IP network and sudden variable delay (spike) often performs both regular and irregular characteristics. Different playout buffer algorithms can have different impacts on the achievement speech quality. It is important to design a playout buffer algorithm which can help achieve an optimum speech quality. In this paper, we investigate to the understanding how network impairments and existing adaptive buffer algorithms affect the speech quality and further to design a modified buffer algorithm to obtain an optimized voice quality. We conduct experiments to existing algorithms and compared their performance under different network conditions with high and low network delay variations. Preliminary results show that the new algorithm can enhance the perceived speech quality in most network conditions and it is more efficient and suitable for real buffer mechanism.*

## 1. INTRODUCTION

Voice over Internet Protocol, which stands for voice over Internet protocol, basically means voice transmitted over a digital network. Well, that isn't technically accurate because the Internet isn't strictly necessary for Voice over Internet Protocol, although it was at first. What is necessary for Voice over Internet Protocol technology is the use of the same protocols that the Internet uses. (A protocol is a set of rules used to allow orderly communication.)Thus, voice over Internet protocol means voice that travels by way of the same protocols used on the Internet. Voice over Internet Protocol is often referred to as IP telephony (IPT) because it uses Internet protocols to make [1] enhanced voice communications possible. The Internet protocols are the basis of IP networking, which supports corporate, private, public, cable, and even wireless networks. Voice over Internet Protocol unites an organization's many locations including mobile workers into a single converged communications network and provides a range of support services and features unequaled in the world of telephony. Technically, IPT refers to telephone calls carried over the organization's local area network (LAN) such as [2] a single building location, a campus-like network, or even a LAN within your home. When IPT crosses from the LAN to the WAN or any other external network, including other LANs operated by the same company at distant locations or the Internet, it becomes Voice over Internet Protocol.

Voice over Internet Protocol is not just about making and receiving telephone calls; it's about a whole new way of communicating. Sure, it includes telephone calls, but there is so much more to the Voice over Internet Protocol telephony picture [3]. Voice over Internet Protocol integrates most if not all other forms of communication. You can even run video conferencing to your desktop [4].

## 2. STANDARDS FOR VOICE OVER INTERNET PROTOCOL

There are multiple signaling standards

H.323, the ITU standard that was published in 1995, started the development of Voice over Internet Protocol products and services. There are four versions available. V.1 is obsolete and has been discontinued in virtually all products. Versions 2, 3 and 4 are used in today's products. These three versions are similar in design and are upwardly compatible. This is the dominant installed signaling protocol for use with hard and soft phones.

The Session Initiation Protocol (SIP) was produced by the IETF as an IP standard. Although SIP is gaining considerable attention, it will not become the dominant installed protocol for a few years. The attractions of SIP are better interoperability among [5] vendors, easier application development, operation that is close to other existing IP protocols, and easier operation through firewalls. It is usually part of hard and soft phones, but it may also be used with gateways. SIP is a completely different design when compared to H.323 [2].MGCP is a protocol used primarily with gateways, although an occasional hard phone may support MGCP.

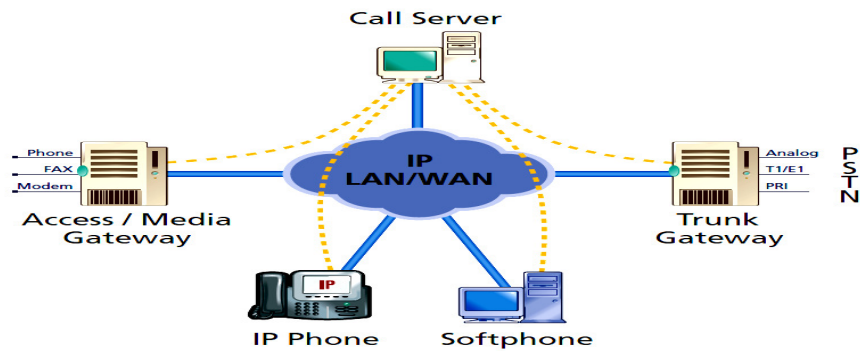

Figure - 1 H.323, SIP and proprietary signalling

MEGACO/H.248, another standard protocol, is a combined effort of the ITU and IETF. It can be used by gateways and server-to server communications. It is not found in hard or soft phones. In addition to the standards, nearly every IP PBX vendor has produced a proprietary signaling protocol. The most commonly found protocols are Cisco's SCCP, or "Skinny," protocol. These proprietary protocols may be variations of the standards or may be uniquely designed.

They each provide the call control found in the standard protocols. An IP PBX vendor usually supports one or more of the standard signaling protocols plus their proprietary protocol. All the signaling protocols follow the same path for control as shown in Figure 1. The H.323, and most proprietary signaling, is carried over TCP, while SIP operates over UDP.

## 3. VOICE OVER INTERNET PROTOCOL AND ASSOCIATED PROTOCOL

Voice over Internet Protocol refers to a type of communication service transported via the Internet. The basic concept of running this application is as follows and presented in Figure 2.

1. Convert the analogue voice signal to digital format.
2. Compress/ translate the converted signal into IP packets.
3. Transmit IP packets to receivers.
4. Reverse the first and second steps at the receiving end.

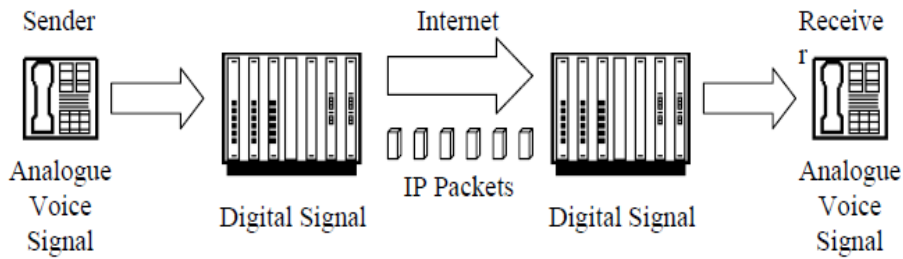

Figure - 2 Basic Concept of Voice over Internet Protocol

To transmit a Voice over Internet Protocol packet, certain protocols or standards are used, as illustrated using the TCP / IP layer distribution in Figure 2. In the application layer, ITU-T H.323 is a popular standard for sending voice and video using IP on the public Internet and within an Intranet. Real-Time Protocol (RTP) [6] and Real-Time Control Protocol (RTCP) are two widely used protocols for managing audio and video signals. RTP is essentially for real-time applications, designed to synchronize different traffic streams in terms of compensation for delay variations and de-sequencing. It however does not ensure the on-time delivery of traffic signals or for recovering lost packets, and does not address the issue of QoS, related to guaranteed bandwidth availability for specific applications. RTCP is another protocol, usually used with RTP. It is based on the periodic transmission of control packets to all participants in the session, using the same distribution mechanism as the data packets. Both RTP and RTCP run on the top of User Datagram Protocol (UDP), which can provide better real-time responsiveness and lower overhead. After compression and digitization of a sample of analogue [7] voice signal, RTP and RTCP information can be packed inside an IP packet with a UDP header to accomplish the requests of transmitting a voice packet.

## 4. HOW VOICE OVER INTERNET PROTOCOL WORK

A Voice over Internet Protocol -based PBX starts up (boots up) like other servers. Once the booting up is complete, the IP phones and gateways can register with the call server. The IP phone and/or gateway must first access a DHCP to obtain an IP address. The DHCP may be part of the data network, or it may be a separate server, or it can be integrated with the call server. Once an address has been assigned, the IP device contacts the call server to register. The call server may have a common set of privileges and restrictions for IP devices or an administrator can make the feature assignments. The call server or another assigned server also adds this device and its phone numbers to the DNS to support directory services [4]. A Permanent H.323 TCP session is established between the Voice over Internet Protocol device and the call server. This is true for most proprietary signaling protocols. SIP uses UDP for the signaling path.

When a user picks up the phone, the dial tone can be generated locally by the phone or by the call server. The IP device then sends one or more packets requesting a connection and the features to be implemented during the connection, such as a conference call. The call server then determines whether the other device is available or busy. If available, the call server contacts the receiving device and instructs both the caller and called devices to establish a peer-to-peer UDP path to [8] carry the RTP speech. The call server becomes dormant during the call until one of the devices terminates the call. The call server then breaks the peer-to-peer connection and records the call event as part of the Call Detail Record (CDR).

Security of Voice over Internet Protocol has become a major concern for Voice over Internet Protocol adopters [9]. Most firewalls do not support Voice over Internet Protocol except through VPN connections. As a default data firewalls will probably prevent the operation of Voice over Internet Protocol by users on the untrusted network when they call devices on the trusted network. Several vendors are now offering encrypted signaling and encrypted speech behind the firewall [10]. This design prevents security problems produced by users of the trusted network. The encryption functions must be implemented in the IP phones, gateways and call servers.

## 5. VOICE OVER INTERNET PROTOCOL SYSTEM AND APPLICATION

### 5.1 VOICE OVER INTERNET PROTOCOL SYSTEM

Figure 3 shows the whole process of voice transferred from host A to B via the representative Voice over Internet Protocol  system, where a number of mechanisms, such as playout buffer and voice/silence detector, are set up in order to achieve the optimized transmission results [11].

### 5.2  PLAYOUT BUFFER

In Voice over Internet Protocol systems, the voice packet must be played out at the receiver site in a timely manner and in the order they were emitted from the sending site. The basic function of a playout buffer is to collect packets and to store them, and then to send a specified number of packets to the next mechanism [12]. Playout buffer is located at the end of Voice over Internet Protocol systems, and the main intention of it is to smooth speech. When variable delays take place, playout buffer can allow some later-arrived packets to [13] be played out, which depends on the set-up of packet playout time, to keep the completeness of speech. However, any packet, arrived later than the playout time, will be simply discarded [14].

The set-up of playout time can be fixed or adaptive, but both need synchronization between sender and receiver due to the changing network delay. Fixed playout time set-up/scheduling is simple, but normally causes a constant delay and cannot follow the change of network delays. Adaptive playout scheduling was introduced to overcome these problems, and is controlled by a corresponding playout buffer algorithm, which can utilize [15] the silence time between two successive voice periods, referred to as talkspurt, to slow down or speed up playout time of each talkspurt.

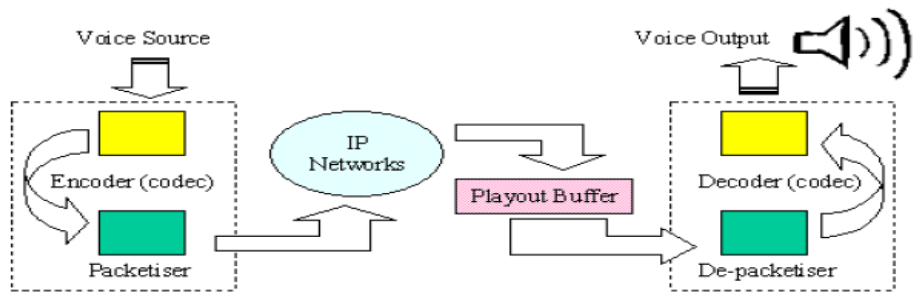

Figure - 3 Voice over Internet Protocol Systems

### 5.3  VOICE AND SILENCE DETECTOR

Human conversation consists of speaking period (talkspurt) and silence gaps, also known as on-off pattern. The function of a voice or silence detector, located on the sender site, is to distinguish these two kinds of period. Voice Activity Detector (VAD) [16] and Silence Detector (SD) are two examples, ubiquitously used in the Voice over Internet Protocol system. The existence of talk spurts and silences allows for silence suppression, [17] where a voice segment is transmitted only if it is detected as a talkspurt. The main benefits of utilizing this kind of mechanism are

1. Allow higher bandwidth utilization through multiplexing.
2. Allow per talkspurt playout time (delay) adjustment.
3. Enable echo suppression based on silence detector output.

## 6. OBJECTIVE METHOD ON MONITORING VOICE QUALITY

There are two broad classes of speech quality metrics subjective and objective. Subjective measures involve humans listening to a live or recorded conversation and assigning a rating to it. In objective measurement methods, the PESQ algorithm provides a more accurate measure of quality. It combines the time-alignment technique for Perceptual Analysis Measurement System (PAMS) with the accurate perceptual modeling of Perceptual Speech Quality Measurement (PSQM), the best features of each technique. PESQ compares the original (reference) signal with the degraded output of the system under test using a perceptual model, as presented in Figure 4.

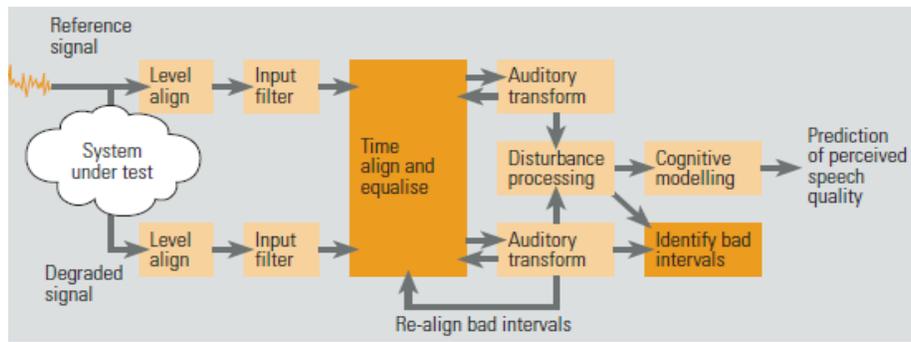

Figure - 4 Basics of PESQ Algorithm

The most eminent result of PESQ is the MOS. It directly expresses the voice quality. The PESQ MOS as defined by the ITU recommendation P.862 ranges from 1.0 up to 4.5, where values close to 4.5 indicate better speech quality, and values close to 1.0 indicate worst speech quality. Since PESQ is an intrusive, end-to-end measurement algorithm and requires a reference speech signal, it can only be to predict one-way listening speech quality. Therefore, the measurement of delay cannot be approached when only using PESQ.

In contrast, E-model is a non-intrusive method, which does not need a reference signal and are appropriate for monitoring live traffic. The quality evaluation of E-model is normally set after jitter buffer and before the concealment equipment compared to PESQ set after the decoding process where the signal has been concealed [18]. Thus, due to the fact of, such as the enhanced technique of error (loss) concealment, E-model may not be well accurate when the packet loss ratio increases. In addition to that, the equations of E-model for assessing loss ratio for different codec's are based on thousands of subjective tests, thus it is really impractical and difficult to develop [19]. For the E-model, it uses the impairment rating scale R to express as the speech quality.

The factor R can be expressed as

$$R = Ro - Is - Ie - Id, eff + A$$

If ignoring the effects of other impairments, the factor R might be simplified as

$$R = Ro - Ie - Id$$

Where Ro is the optimum quality value (the default value for Ro can be set to 93.2), Ie is known as the equipment impairment factor and accounts for impairments due to non-linear codec and packet loss. Id accounts for echo and delay [19].

Moreover, if related the factor R to the MOS, as mentioned in ITU-T Recommendation, if given the R the value, the corresponding MOS (PESQ) can be obtained as

$$MOS = 1 \qquad \text{for } R <= 0$$
$$MOS = 1 + 0.035R + R(R-60)\ (100-R)\ x7x10^{-6} \text{ for } 0 < R < 100$$
$$MOS = 4.5 \qquad \text{for } R >= 100$$

For the equation, if the R value is smaller than 6.5, then the MOS value will be below 1. Thus the R value is usually restricted to the range 6.5 ~ 100.

# 7. PROPOSED METHOD ON EVALUATING VOICE QUALITY

The PESQ or E-model has its own weaknesses at the same time has its own merits. Thus it is better to utilize both advantages to enhance the accuracy and efficiency of evaluation of voice quality, especially of conversational speech quality. In other words, the precise assessment on the impact of packet loss and the evaluation of packet delay can be achieved by using PESQ algorithm and E-model [19]. The Figure 5 presents the concept of this methodology. The listening MOS (PESQ) is firstly obtained using PESQ algorithm. Then as described before, the R value of E-model can be converted from the [20] MOS equation. Subsequently, the Ie value is deduced by R value. Finally, the conversational MOS can be derived from R value, combined of both Ie and Id value.

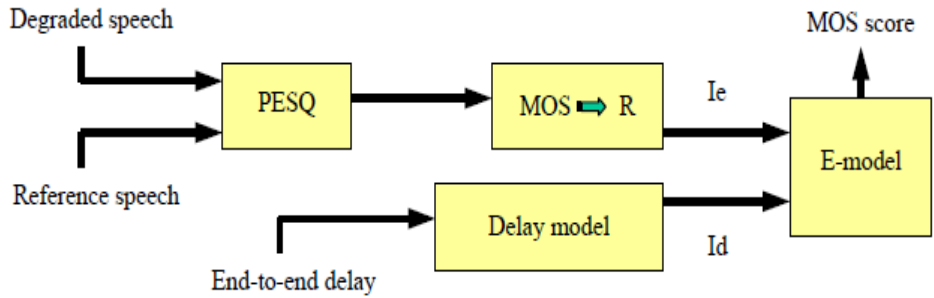

Figure - 5 Speech Quality Measurements

For ease of mapping from MOS to R value, a simplified $3^{rd}$ order polynomial fitting as presented below, to achieve the conversion.

R = 3.026 MOS3 – 25.314 MOS2 + 87.060 MOS - 57.336

If only considering the delay impairment factor or the equipment impairment Ie, we can correspondingly obtain the Ie and Id values. For the Id value, it can be calculated as

Id = 0.024 Ta + 0.11(Ta – 177.3) H (Ta – 177.3)

Where

        H(X)   =   0            if X < 0
        H(X)   =   1            if X >= 0

Ta represents an absolute delay (playout delay).

For the Ie value, as using PESQ algorithm to derive this value [21]. The equation of Ie can be obtained from the curve of MOS (PESQ) vs. random packet loss ratio as

Ie = 20.06 x ln (1+ 0.1024 x loss ratio) + 25.63

This method of evaluating conversational speech quality can also be extended to other speech codes. It can be easily used to monitor and predict conversational speech quality of the network impairments non-intrusively [22].

# 8. ADAPTIVE BUFFER ALGORITHM

## 8. 1 General Estimate of Playout Delay

### 8. 1. 1 Exponential-Average (Exp-Avg)

This algorithm uses the mean $d_i$ and variance $v_i$ to estimate the playout delay $p_i$ of ith arriving packet as

$$p_i = d_i + 4v_i$$

Where di and vi are derived as,

$$d_i = \alpha d_{i-1} + (1-\alpha) \, ni$$
$$v_i = \alpha v_{i-1} + (1-\alpha) \, |d_i - n_i|$$

Where $n_i$ denotes the one-way delay of the ith packet, and the value of $\alpha$ is defined as 0.998002.

### 8. 1. 2 Min-Delay (Min-D)

The purpose of this algorithm is to minimize the delays. It uses the minimum delay of all packets received by the current talkspurt as $d_i$ to predict the next talk spurt's playout delay. Let $S_i$ be the set of all packets received during the talkspurt. The $d_i$ is computed as

$$d_i = \min_{j \in si} \{n_i\}$$

### 8. 1. 3 Spike Detection

This algorithm contains a spike detection mechanism. When a spike is detected, [23] the algorithm changes in SPIKE mode, the delay estimate tracks the delays closely. If it is not in the SPIKE mode, then the concept of this algorithm is same as the first algorithm, but except the value of $\alpha$ is set to 0.875.The Spike-Det Algorithm as shown below.

$n_i$ :  ith packet network delay

IF (mode == NORMAL)
    {
        IF (abs ($n_i - n_{i-1}$) > abs ($v_i$) ∗ 2 + 800)
          {
               var = 0;
                mode = SPIKE;
          }
        ELSE
          {
           var = var/2 + fabs (2 $n_i - n_{i-1} - n_{i-2}$) / 8;
            IF (var<=63)
              {
                   mode=NORMAL;
              $n_{i-2} = n_{i-1}$;
              $n_{i-1} = n_i$;
                return;
                }
          }
        }

IF (mode == NORMAL)

$\qquad$ $d_i = \alpha d_{i-1} + (1-\alpha)n_i;$

$\quad$ ELSE

$\qquad$ $d_i = d_{i-1} + n_i - n_{i-1:}$

$\qquad\quad$ $v_i = \alpha v_{i-1} + (1-\alpha)|d_i - n_i|;$

$\qquad\quad$ $n_{i-2} = n_{i-1};$

$\qquad\quad$ $n_{i-1} = n_i;$

return;

## 9. PROPOSED TO THE DESIGEN OF BUFFER ALGORITH

After studying the concepts and problems of existing adaptive playout buffer algorithms, to sum up, a good buffer algorithm should contain all the following points:

1. Should relate to the view of perceived speech quality.
2. Can properly follow the change of network delay.
3. Can present a good trade-off when variable delays are small.
4. Should have a nimble spike detection mechanism.
5. Can predict a proper playout delay (time) when a spike occurs.
6. Can decrease the playout delay in an appropriate speed after the happening of a spike
7. Can be suitable for any network circumstance.
8. Should have a high calculation(estimate the playout delay) speed for every received packet.
9. Can/May well perform and/or collaborate with different set-up of Voice over Internet Protocol mechanisms.

## 10. PROPOSED ALGORITHEM

Calculating Variance for last **n** packets

$V_i = V_{i-1} + (\Delta d_0{}^2 - \Delta d_i{}^2) / n$

Where,

$\qquad$ $V_{i-1}$ → variance of previous talkspurt

$\qquad$ $\Delta d_0$ → delay deviation of nth last packet

$\qquad$ $\Delta d_i$ → delay deviation of current packet

Along with this, a variable sign_generator is kept which is changed according to the following algorithm.

If (delay of the last packet > threshold * delay of current packet && talkspurt is not beginning)

$\qquad\qquad$ sign_generator = -1;

$\quad$ else

$\qquad\qquad$ sign_generator = 1;

The threshold variable depends on the network type. To decide threshold at runtime, we keep a check on the maximum and minimum playout delays received during each talkspurt. The max and min variables are reset at the beginning of each talkspurt [24]. The threshold variable is calculated as

If (max > 3.0 * min)

threshold = 1.75; (for high-jitter networks)

else if (max > 1.5 * min)

threshold = 1.5; (for medium-jitter networks)
else
threshold = 1.1; (for stable networks)

The predicted playout delay for normal mode then becomes,

$d_i = d_{i-1} + \sqrt{V_i} *$ sign_generator

In spike mode, the algorithm just takes [25] playout delay to be equal to packet delay. This algorithm can be translated into a pseudo-code format as shown below.

```
If (mode == NORMAL)
{
     Vi = Vi-1 + (Δd0² - Δdi²) / n;
          If ( (di-1 > threshold * di)  &&  (di < head * pk) )
               {
                    sign_gen = -1;
               }
                else
                {
                     sign_gen = 1;
                }
          di = di-1 + √Vi * sign_gen;
}
   else

{
     if ( mode == SPIKE )
          {
               if (dik <= tail * old_d)
             {
                     mode = NORMAL;
             }
          }
     else
          {
          if (dik > head* pk)
               {
                    mode = SPIKE;
                    old_d = pk;
               }
          }
      }
```

## 11. COMPARISON WITH EXISTING ALGORITHEM

All the four algorithms were run at identical conditions and the delays were simulated for ten thousand (10,000) packets transmitted over three different types of networks.

      A.) Stable Delay Network
      B.) Medium Jitter Network
      C.) High Jitter Network.

Since the Packet delay is normally estimated to be higher in variable delay and lower when delay changes steadily, the threshold need to be suitable for these changes, as for trace PU2CU, we can set the threshold to be lower due to the steady network delay, and for the trace PU2BU, set to be higher owing to the fluctuate network delay. Table compares the Average Packet Delay, Packet Loss ratio and Mean Opinion Score (MOS) of each algorithm in three traces. The first one traces as BU2PU and PU2BU have more variable network circumstances, and the last two traces are more steady networks. From the results in table the Exp-Avg algorithm can perform well in the latter two traces, but cause huge Packet delays in the former two traces. Thus it is not recommended to use in such variable networks. For other algorithms, some work better in stable networks and some in uneven networks. However, the Suggested algorithm performs well in both cases. It can automatically adapt different network circumstances. As for the variable network, it allows slightly increase of packet loss to obtain the considerable decrease of Packet delay. In stable networks, it ensures the smaller packet loss ratio, simultaneously, achieves the lower Packet delay. In other words, it can accomplish a good trade-off as between Packet delay and packet loss ratio.

### A. Stable Delay Network

Base Delay = 50ms

| Algorithm Used | Average Packet Delay (ms) | Loss Percentage (%) | Mean Opinion Score (MOS) |
|---|---|---|---|
| Exp-Avg | 52.83 | 3.21 | 3.13 |
| Min-D | 51.89 | 7.29 | 2.84 |
| Spike-Det | 52.75 | 4.58 | 3.03 |
| Suggested | 54.73 | 0.71 | 3.35 |

Table -1 Comparison (Stable Network)

Base Delay = 100ms

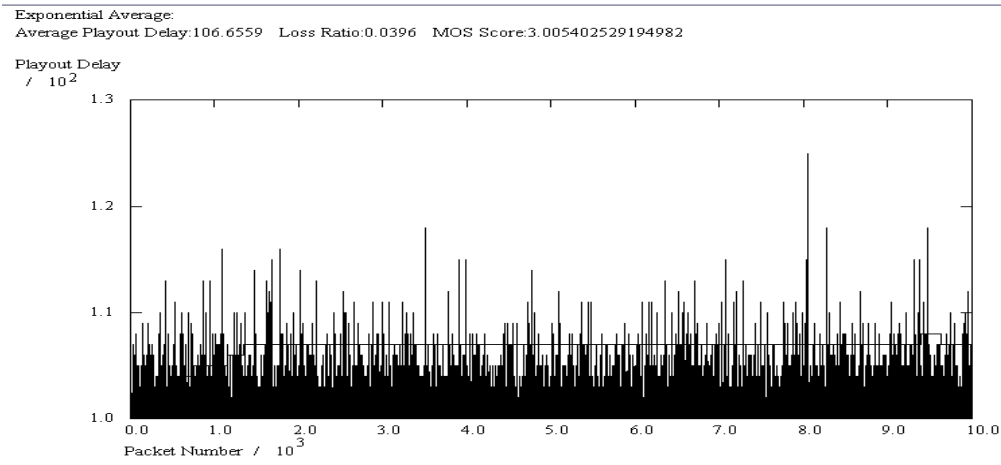

Figure - 6  Exp-Avg (Stable Network)

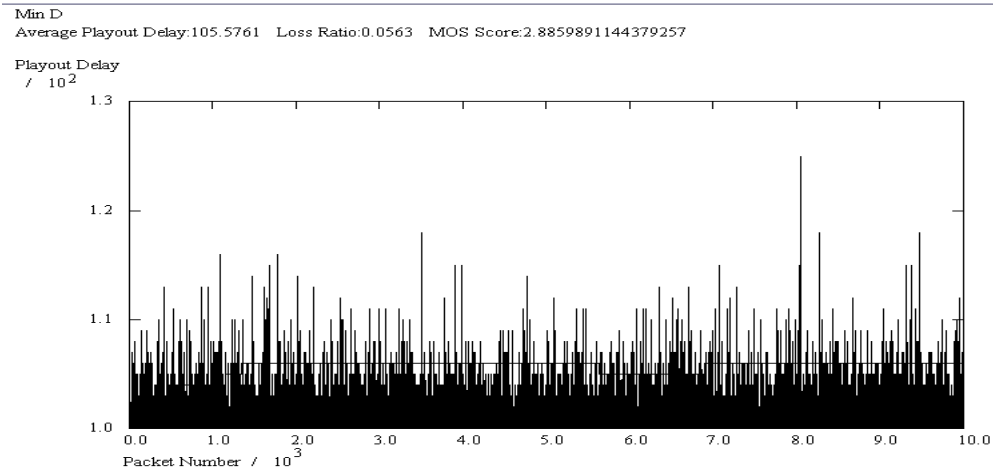

Figure - 7  Min-D (Stable Network)

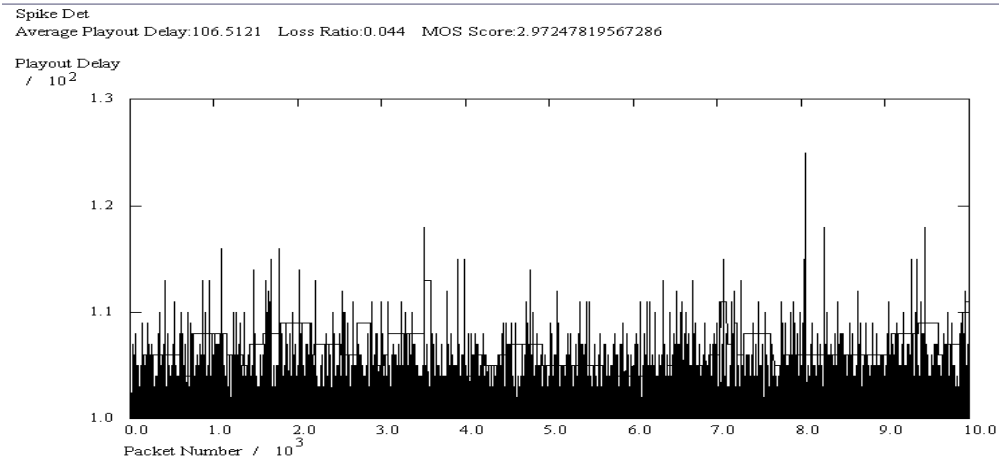

Figure - 8 Spike-Det (Stable Network)

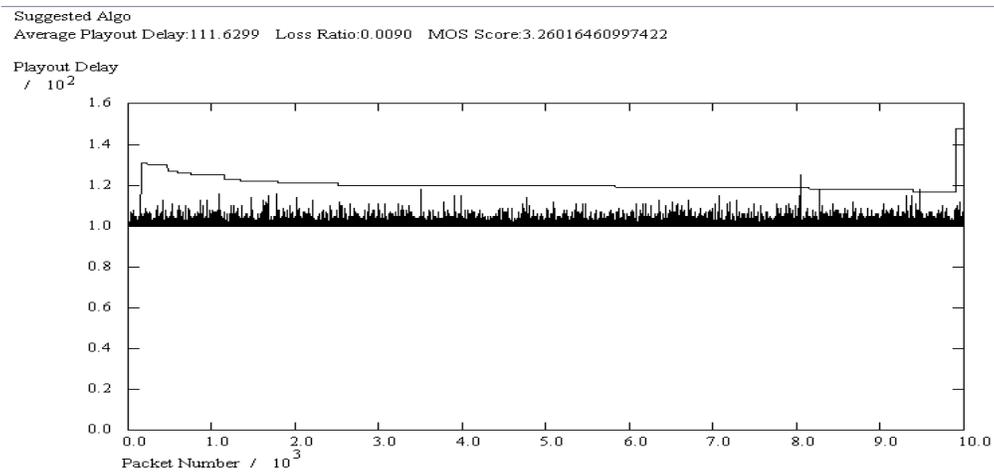

Figure - 9 Suggested (Stable Network)

## B. Medium Jitter Network

Base Delay = 50ms

| Algorithm Used | Average Packet Delay (ms) | Loss Percentage (%) | Mean Opinion Score (MOS) |
|---|---|---|---|
| Exp-Avg | 99.8 | 4.67 | 2.96 |
| Min-D | 95.31 | 5.55 | 2.9 |
| Spike-Det | 190.33 | 1.86 | 2.99 |
| Suggested | 104.46 | 1.32 | 3.23 |

Table- 2 Comparisons (Medium Jitter)

## Base Delay = 50ms

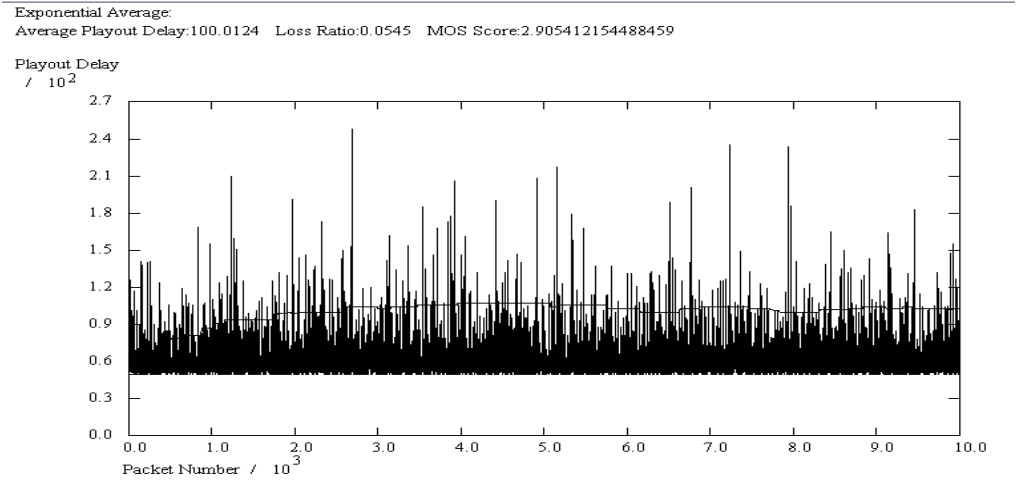

Figure - 10 Exp-Avg (Medium Jitter)

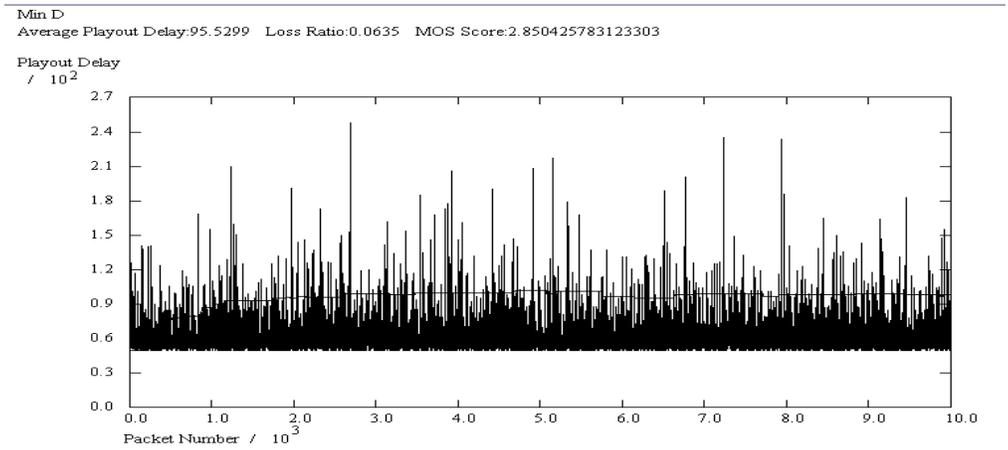

Figure - 11 Min-D (Medium Jitter)

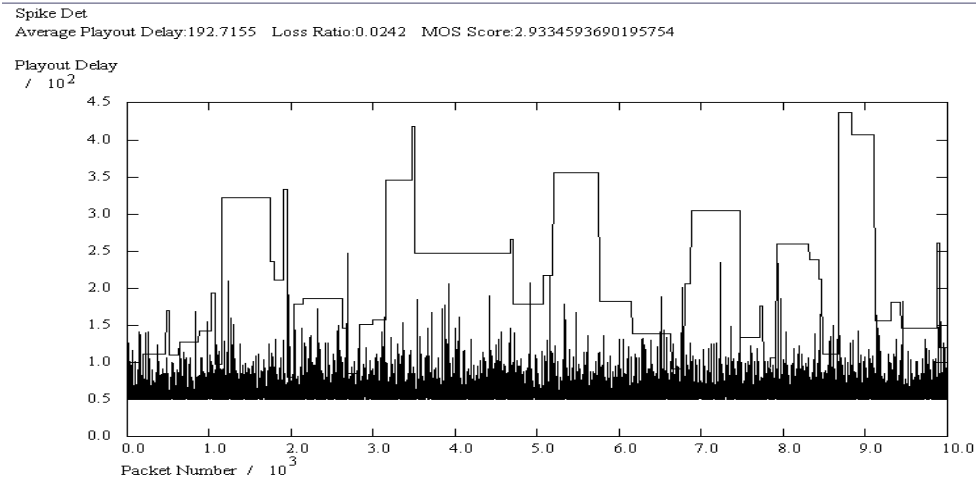

Figure - 12 Spike-Det (Medium Jitter)

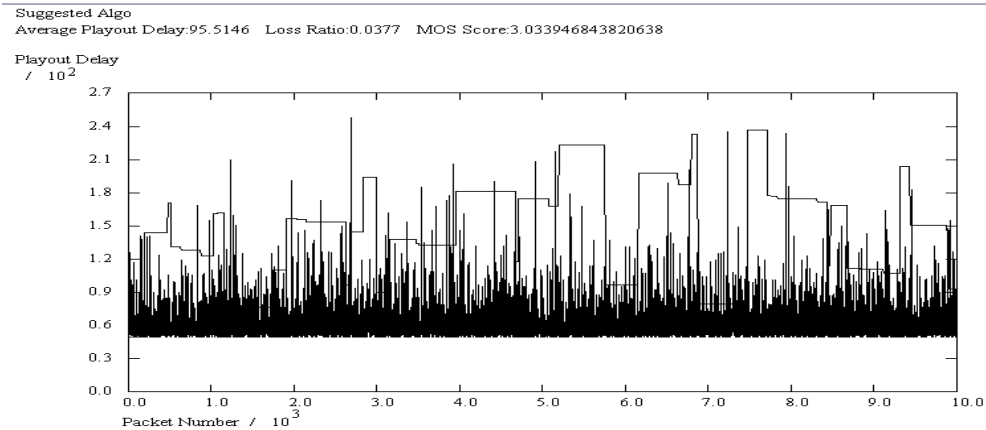

Figure - 13 Suggested (Medium Jitter)

The Stable Delay, Medium Jitter and high Jitter network these delay packets are then randomly divided into talkspurts of varying lengths. The proposed Algorithm is then used to generate the playout delays for the above delay packets. The Packet delays are recorded and then used to calculate the Average Packet delay, packet loss ratio and MOS scores for each algorithm. The variation of playout delay along with the packet delay is also shown graphically. In the proposed algorithm table 1 shows the Stable Delay Network selected, the higher MOS can be obtained. The performance of if we increase the value of the head, the average packet delay goes up , at the same time the packet Loss ratio goes down, the MOS scores are higher . It seems to change the sets of threshold of these algorithms will considerably influence the speech quality. Another medium jitter and high jitter network selected the performance of if we increase the MOS scores than at the same time the packet Loss ratio goes down result shown table 2 and table 3. We compare our suggested algorithm with each of the other existing algorithms for each kind of network as tabulated.

## C. High Jitter Network

Base Delay = 50ms

| Algorithm Used | Average Packet Delay (ms) | Loss Percentage (%) | Mean Opinion Score (MOS) |
|---|---|---|---|
| Exp-Avg | 150.56 | 4.77 | 2.89 |
| Min-D | 138.56 | 5.9 | 2.82 |
| Spike-Det | 471.95 | 1 | 1.31 |
| Suggested | 114.69 | 4.28 | 2.97 |

Table - 3 Comparisons (High Jitter)

Base Delay = 50ms

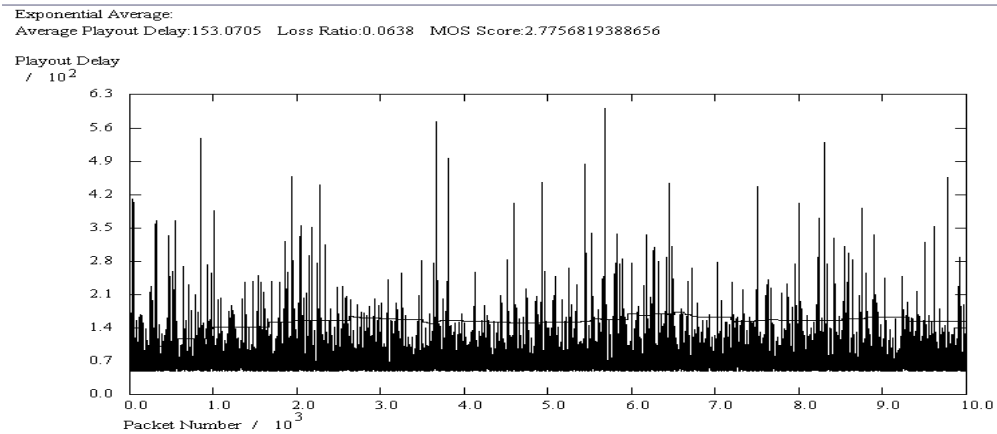

Figure - 14 Exp-Avg (High Jitter)

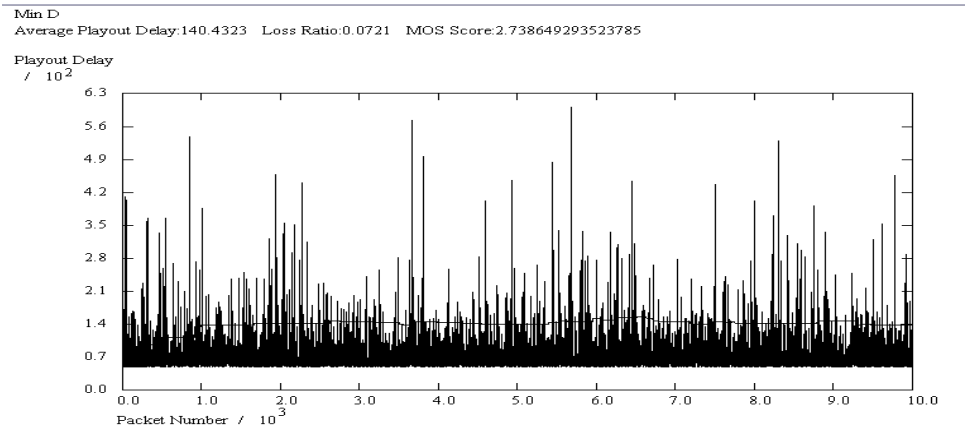

Figure - 15 Min-D (High Jitter)

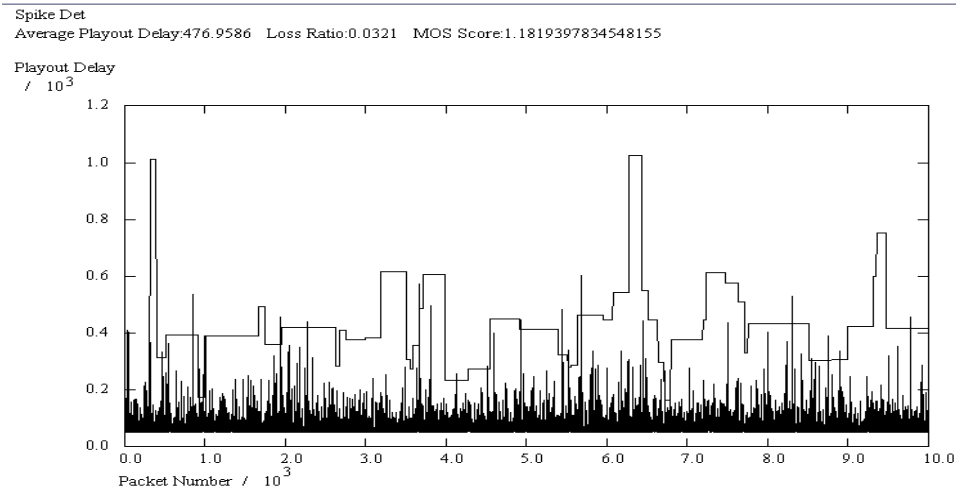

Figure - 16 Spike-Det (High Jitter)

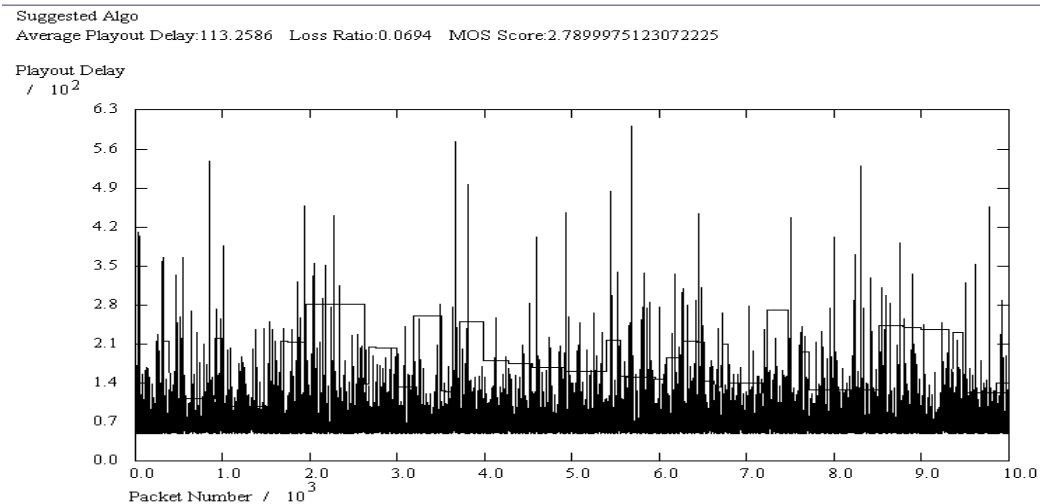

Figure - 17 Suggested (High Jitter)

The proposed algorithm can follow the tendency of the change of the network delay, and results in higher MOS scores during each time interval, compared to others, which may not perform well during some periods. The comparison of simulation results shows that our new algorithm can enhance the perceived speech quality in most network circumstances. it can follow the variation of network delay, and can give a proper Packet delay (time) for each talkspurt. Thus, our new buffer algorithm is more efficient and is suitable for any real buffer mechanism.

## 11.CONCLUSIONS

Voice over Internet Protocol (VoIP) is one of the most important technologies in the world of communication. Around 20 years of research on VoIP, some Quality of Service (QoS) problems of VoIP is still remaining.An end-to-end packet delay in the Internet is an important performance parameter, because it heavily affects the quality of various applications including real-time and data applications. We are proposing a new playout buffer algorithm. The conclusion of this paper depicts that the Superior cost benefit and suitable quality are two important points to make the voice over IP networks attractive. As with other communication means, quality is the core portion of QoS for Voice over Internet Protocol applications. Such network parameters as delay, delay variation (jitter), packet loss etc., Can have a significant influence on speech quality. Playout buffers, which can be controlled by buffer algorithms, are typically utilized to compensate for jitters. The research on buffer algorithms shows that different buffer algorithms/schemes can result into different levels of speech quality. The analysis and experimental results present the performance of existing playout buffer algorithms, and shows that they can just be acceptable in certain types of network circumstances. Thus the design of a buffer algorithm should be based on diverse network features and the view of speech quality. The comparison of corresponding simulation results shows that our new algorithm, designed based on the above-mentioned two points, can further enhance the speech quality, and is more efficient as well as suitable for any real buffer mechanism.